\documentclass[a4paper, 11pt, fleqn]{article}
\usepackage[utf8]{inputenc}
\usepackage[english]{babel}
\usepackage{longtable}
\usepackage{graphicx}
\usepackage{amssymb,amsmath,epsfig,array}

\begin{document}

\selectlanguage{english}

\title{IRAS 22272+5435 (V354 Lac): Multicolor Photometry of a Variable Carbon-rich post-AGB Star and a Dust Formation Episode}

\author{ N.~P.~Ikonnikova$^{1}$\footnote{e-mail: ikonnikova@sai.msu.ru}, V.~I.~Shenavrin$^1$, G.~V.~Komissarova$^1$, M.~A.~Burlak$^1$}

\date{\it{$^1$Sternberg Astronomical Institute, Lomonosov Moscow State University, Moscow, Russia}}

\renewcommand{\abstractname}{ }

\maketitle

\begin{abstract}

We present new $UBV$ photometric observations (2009–2024) of the carbon-rich post-AGB star IRAS 22272+5435 (V354 Lac), combined with our data spanning over 30 years (1991–2024). The star shows pulsations with two closely spaced periods, 127 and 132\,d, which cause brightness variations of variable amplitude. We also detect low-amplitude oscillations with a 661-day period, of uncertain origin. Color analysis reveals both temperature variations during pulsations and spectral anomalies. For the first time, we analyze long-term $JHKLM$ photometry (1994–2024), showing increased $KLM$-band brightness from 1996 to 2004 -- likely due to dust emission following a sudden mass ejection. The absence of strong optical variability suggests either a large angle between the ejection direction and the line of sight or a low optical depth of the ejected material. We discuss possible mechanisms and observational signatures of this dust-formation event.

{\it Keywords}: post-AGB supergiants, photometric observations, semiregular variable stars, evolution, dust shells

\end{abstract}
\bigskip
\bigskip

\section*{INTRODUCTION}

Stars with initial masses of $M_{\text{ZAMS}} = 0.8-8.0 M_{\odot}$ undergo hydrogen and helium shell burning around a degenerate core during the asymptotic giant branch (AGB) phase. A key feature of this stage is intense mass loss via stellar winds up to 10$^{-4}$ $M_{\odot}$/yr -- which eventually leads to the ejection of the outer envelope. In the subsequent post-AGB phase, these stars retain extended atmospheres around their cores, making their spectra resemble those of supergiants. They are also surrounded by gas/dust shells composed of material expelled during thermal pulses on the AGB. As a result, a defining characteristic of post-AGB stars is the presence of a significant infrared (IR) excess. According to the catalogue by Szczerba et al. (2007), about 300 post-AGB stars and related objects are currently known.

IRAS 22272+5435 was one of the first IR sources that initiated comprehensive studies of post-AGB stars. It was first mentioned by Pottasch and Parthasarathy (1988), who reported the detection of an IR excess in ten high-luminosity stars of spectral classes F to G, indicating the presence of circumstellar dust shells. The energy distribution in the far-IR, as well as the luminosities and dust shell masses of these objects, were found to be similar to those of planetary nebulae. Based on this evidence, Pottasch and Parthasarathy (1988) concluded that these stars, including IRAS 22272+5435, belong to the AGB or post-AGB evolutionary phase. The IR source IRAS 22272+5435 was later identified with the bright star BD+54$^{\circ}$2787 = HD~235858 = SAO~034504 = V354~Lac ($22^{\text{h}}29^{\text{m}}10.4^{\text{s}}, +54^{\circ}51'06''$ (2000)).

Van der Veen et al. (1989) obtained the first near-IR observations of the star, constructed its spectral energy distribution (SED), and classified it into group IVa -- whose members show a double-peaked SED corresponding to emission from the stellar photosphere and a detached dust shell. Hrivnak and Kwok (1991) obtained an optical spectrum of IRAS 22272+5435 and detected strong molecular bands of C$_{3}$ and C$_{2}$. Based on this finding, along with results from earlier studies cited in their work, they concluded that IRAS 22272+5435 belongs to the class of carbon-rich post-AGB stars. IRAS 22272+5435 is one of the few post-AGB objects whose IR spectra show distinct emission features at 21\,$\mu$m (Kwok et al., 1989) and 30\,$\mu$m (Omont et al., 1995). The nature of the carriers responsible for these features remains unknown, but various hypotheses have been proposed and widely discussed (e.g., Cox, 1990; Goebel, 1993; Posch et al., 2004; Zhang et al., 2009).

Later, Za\v{c}s et al. (1995) determined the chemical composition of IRAS 22272+5435 for the first time and found it typical of carbon-rich post-AGB stars. They reported a deficiency in iron ([Fe/H] = –0.49) and an enhancement of $\alpha$-elements relative to solar abundances. These characteristics clearly indicate an advanced evolutionary stage. The atmospheric parameters derived by Za\v{c}s et al. (1995) -- effective temperature $T_{\text{eff}} = 5600$\,K, surface gravity $\log g=0.5$, and microturbulent velocity $\xi_{t}=3.7$\,km/s -- are consistent with later estimates from Za\v{c}s et al. (1999), Reddy et al. (2002), Klochkova et al. (2009), and De Smedt et al. (2016). The star’s atmospheric composition has been studied multiple times. Klochkova et al. (2009) measured the abundances of 22 elements and found significant overabundances of s-process elements (Ba, La, Ce, Nd). In one of the most recent studies, De Smedt et al. (2016) derived abundances for 32 elements in the atmosphere of IRAS 22272+5435. According to their results, the C/O ratio is $1.46\pm0.26$, and [Fe/H] =$-0.77\pm0.12$.

Images of IRAS 22272+5435 clearly reveal a surrounding circumstellar shell. Mid-IR studies by Meixner et al. (1997) and Dayal et al. (1998) showed a significant density contrast between the equatorial and polar regions, suggesting the presence of an inclined dust torus or disk. Observations with the Hubble Space Telescope detected a compact, axisymmetric nebula measuring approximately $3''\times3''$, identified as a proto-planetary nebula (Ueta et al., 2000). The circumstellar envelope is also observed in molecular line emission from CO (Zuckerman et al., 1986; Bujarrabal et al., 2001; Hrivnak and Bieging, 2005; Fong et al., 2006; Nakashima et al., 2012), as well as HCN, CS, and CN (Lindqvist et al., 1988). The expansion velocity of the shell is estimated to be between 9.6 and 13.3 km/s.

An important characteristic of V354 Lac is its photometric variability, first detected by Strohmeier and Knigge (1960) using photographic plates. The star is classified in the General Catalogue of Variable Stars (Samus et al., 2017) as Lb -- a slow irregular variable. Intensive studies of its variability began after its identification with an IR source. Hrivnak and Kwok (1991) confirmed the star's variability with an amplitude of $0.^m8$ in the $V$-band and also detected near-IR variations with an amplitude of about $0.^m2$. Arkhipova et al. (1993, 2000, 2009) refined the nature of the variability and investigated its color characteristics. Based on observations from 1994 to 1996, Hrivnak and Lu (2000) determined a period of $P$=127\,d and attributed the variability to stellar pulsations. In subsequent studies (Hrivnak et al., 2010, 2013, 2017, 2022), the periods were further refined, and the relationships between light curves, color indices, and radial velocities were explored. 

In addition to photometric variability, spectral variability has also been detected in V354 Lac. Spectral monitoring by Za\v{c}s et al. (2009) revealed regular radial velocity changes with an amplitude of about 10\,km/s and a period of 131.2\,d. The authors also observed significant variations in the intensities of C$_2$ and CN lines, variable H$\alpha$ emission, and the appearance of carbon emission features. Further observations by Za\v{c}s et al. (2016) between 2010 and 2011 showed pronounced variations in molecular and atomic line intensities that correlated with the pulsation phase. Recently, Za\v{c}s et al. (2021) reported short-term (less than two weeks) spectral changes, including notable variations in the intensity and position of the CN(1, 0) band near the star’s brightness maximum. These variations are attributed to large-scale convective motions and shock waves propagating through the extended atmosphere of this post-AGB supergiant.

Although V354 Lac is one of the most extensively studied post-AGB stars, investigations into its photometric and spectral variability remain highly relevant. This is because the star is currently undergoing the fastest phase of its evolution -- the transitional stage from the AGB to the planetary nebula phase. During this evolutionary phase, a decrease in the pulsation period is expected as the star’s effective temperature rises and it moves horizontally across the Hertzsprung–Russell diagram toward the domain of planetary nebula nuclei. In addition, the rate and mechanisms of mass loss are still poorly understood, and these can only be constrained through continued observations. Another important aspect is the detection of binary systems among post-AGB stars, as binarity significantly influences stellar evolution. The hypothesis of binarity for V354 Lac proposed by Hrivnak et al. (2022) requires further confirmation through new observational data.

In this work, we present results from a new stage of $UBV$ observations. For the first time, we provide and analyze a long-term $JHKLM$ dataset spanning 1994-2024, which allowed us to trace the dust ejection event that occurred between 1996 and 2004.

\section*{$UBV$-PHOTOMETRY}

We began $UBV$ observations of V354 Lac in 1990 using a photon-counting photometer (Lyutyi, 1971) mounted on the 60-cm Zeiss reflector at the Crimean Astronomical Station (CAS) of the Sternberg Astronomical Institute, Lomonosov Moscow State University (SAI MSU). The results from 1990 to 2008 were published by Arkhipova et al. (1993, 2000, 2009). After 2008, observations of V354 Lac were continued with the same telescope and photometer. We used HD~235865 (Sp=M2Ib–II), with magnitudes $U=12.^m79$, $B=10.^m45$, $V=8.^m54$ (accurate to $0.^m01-0.^m02$), as the comparison star, following our earlier studies. We found no brightness variations in HD~235865 exceeding the measurement errors when comparing it with check stars.

From July 1, 2009 to December 20, 2024, we obtained 207 $UBV$ photometric measurements of V354~Lac. The average observational errors were about $0.^{m}01$ in the $B$ and $V$ bands and $0.^{m}02$ in the $U$ band. The results of the $UBV$ photometry, transformed to the Johnson system using the calibration equations from Arkhipova et al. (2016), are presented in Table~1\footnote{\text{Table 1 is published only in electronic form and accessible via http://vizier.u-strasbg.fr/cats/J.PAZh.htx}} .

\section*{IR PHOTOMETRY}

The IR photometry of V354~Lac began on June 25, 1991 and has been carried out using a photometer with an indium antimonide (InSb) photovoltaic detector cooled by liquid nitrogen, mounted at the Cassegrain focus of the 1.25-m telescope of the Crimean Astronomical Station (CAS SAI MSU). The diameter of the entrance aperture was $\approx 12^{\prime\prime}$, and the spatial separation of the beams during modulation was $\approx 30^{\prime\prime}$ along the east–west direction. The photometric standard used was the star BS8538 from the Johnson et al. (1966) catalog. Initial results of the IR photometry of V354~Lac were announced but not explicitly presented in Shenavrin et al. (2011). A summary of all observations obtained between 1991 and 2024 is given in Table~2\footnote{\text{Table 2 is published only in electronic form and available via http://vizier.u-strasbg.fr/cats/J.PAZh.htx}}. The average observational errors are: $\Delta J=0.^m01$, $\Delta H=0.^m01$, $\Delta K=0.^m01$, $\Delta L=0.^m01$, $\Delta M=0.^m03$.

\section*{PHOTOMETRIC ANALYSIS}

\subsection*{\bf Optical range}

\begin{figure*}
    \includegraphics[scale=1.5]{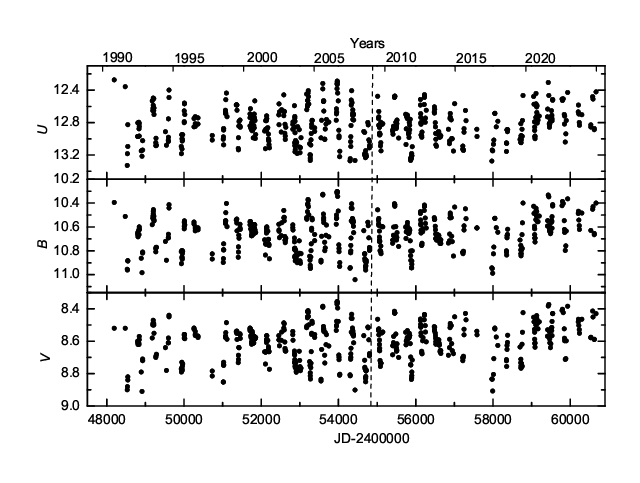}
    \caption{Light curves of V354~Lac in the $UBV$ bands from 1991 to 2024. Observations obtained during 2009-2024 are shown to the right of the vertical dashed line}
    \label{fig:LC}
\end{figure*}

The light curves of V354~Lac in the $UBV$ bands from 1991 to 2024 are shown in Fig.~\ref{fig:LC}. The star exhibits quasi-periodic brightness variations with variable amplitudes reaching up to: $\Delta U=0.^{m}96$, $\Delta B=0.^{m}67$ and $\Delta V=0.^{m}54$. The amplitude changes are due to the superposition of oscillations with closely spaced periods, as previously noted by Arkhipova et al. (2009) and Hrivnak et al. (2010, 2013, 2017, 2022).

%----------------Fig. 1 -----------------------------

%--------------------------------------------------

\begin{figure*}
    \includegraphics[scale=1.5]{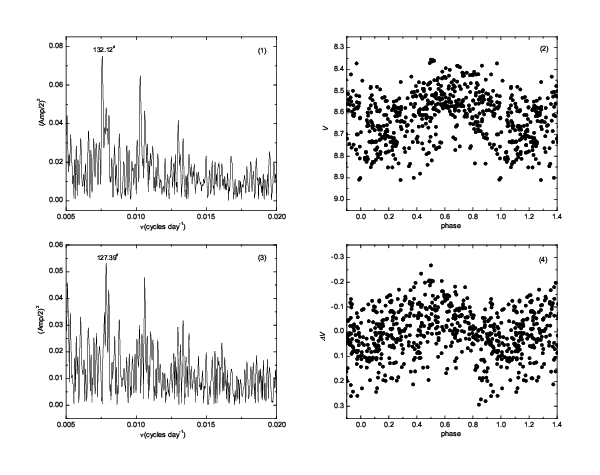}
    \caption{Top row: periodogram (a) and the convolution of the $V$-band data with a period of 132.12\,d (b), based on the full photometric dataset from 1991 to 2024. Bottom row: periodogram (c) and phase curve folded with a period of 127.39\,d (d), derived from the residual light curve after prewhitening.}
    \label{fig:phaseV}
\end{figure*}

For the entire $UBV$ photometric dataset covering 1991-2024, we carried out a period search using V.P.~Goranskij’s WINEFK program\footnote{http://vgoranskij.net/software/} , which implements the method of Diming (1975).We detected two significant periods: 132.12\,d and 127.39\,d. Fig.~\ref{fig:phaseV} presents the periodogram for the $V$-band data along with the convolution corresponding to the dominant period of 132.12\,d. Also shown are the periodogram of the residual light curve after removing the 132.12-d signal and the phase curve computed for the 127.39-d period.

%----------------Fig. 2 -----------------------------

%--------------------------------------------------

A period of 661\,d is clearly detected in the full $UBV$ photometric dataset across the 150–800\,d range. This value closely matches the $\sim$1.8-yr period reported by Hrivnak et al. (2022), based on data from the ASAS-SN survey and observations at the Valparaiso University Observatory (VUO). The authors proposed that this period could be of orbital origin, suggesting the presence of a binary companion; however, further independent observations are needed to verify this hypothesis.

\begin{figure*}
    \includegraphics[scale=1.5]{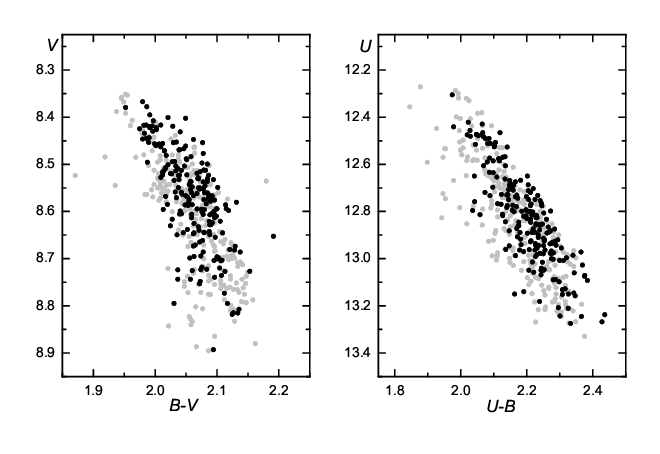}
    \caption{Color-magnitude diagrams based on observations from 1991 to 2008 (grey points) and new data from 2009 to 2024 (black points).}
    \label{fig:CB}
\end{figure*}

Let us examine how the star’s color changes with its brightness. Fig.~\ref{fig:CB} shows the dependence of the $B-V$ and $U-B$ color indices on the brightness in the $V$ and $U$ bands, respectively. In both cases, a clear correlation is observed: as the brightness increases, the star becomes bluer -- the color indices decrease -- indicating an increase in stellar temperature with rising luminosity. This behavior is typical of temperature pulsations. Similar trends are seen in the $(V-R)-R$ diagrams for V354~Lac presented by Hrivnak et al. (2013).

%----------------Fig. 3 -----------------------------

%--------------------------------------------------

\begin{figure*}
    \centering
    \includegraphics[scale=1.5]{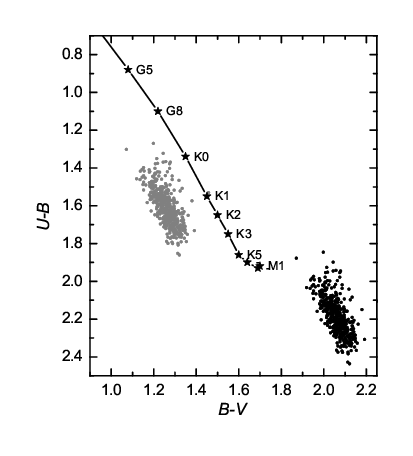}
    \caption{The $(B-V)-(U-B)$ two-color diagram based on observations from 1991 to 2024. Black dots represent observed; grey dots are dereddened values using $E(B-V)=0.^m8$.}
    \label{fig:2col}
\end{figure*}

The two-color diagram $(B-V)-(U-B)$ (Fig.~\ref{fig:2col}) reveals a clear correlation between the color indices: as $B-V$ decreases, $U-B$ also decreases. Comparison of the star’s colors with intrinsic values of normal supergiants (Strai\v zys, 1977) shows that reddening corrections are insufficient to reconcile the photometric data with spectral and empirical models. This discrepancy is due to the peculiar spectrum of V354~Lac, which contains numerous absorption lines of elements brought to the surface during the AGB phase, as well as molecular bands formed in the upper atmospheric layers and circumstellar envelope. These spectral features result in non-standard color indices, making direct comparisons with those of normal supergiants impractical. Ueta et al. (2001) noted that the best-fitting model spectrum for IRAS~22272+5435 corresponds to an interstellar extinction of $A_V = 2.^{m}5$, or a color excess of $E(B-V) = 0.8$. In Fig.~\ref{fig:2col}, the dereddened colors based on this value lie slightly below the sequence of intrinsic color indices of normal supergiants. Variations in these colors during pulsation occur along this sequence and are associated with changes in stellar temperature. During pulsation, the star’s effective temperature varies by approximately 300--400\,K, corresponding to a spectral class shift of 2-3 subclasses. According to estimates by Za\v cs et al. (2016), the change in color temperature derived from variations in the $V-R_C$ color index amounts to about 500\,K.
  
%----------------Fig. 4 -----------------------------

%--------------------------------------------------

\subsection*{\bf IR range}

\begin{figure*}
    \includegraphics[scale=1.5]{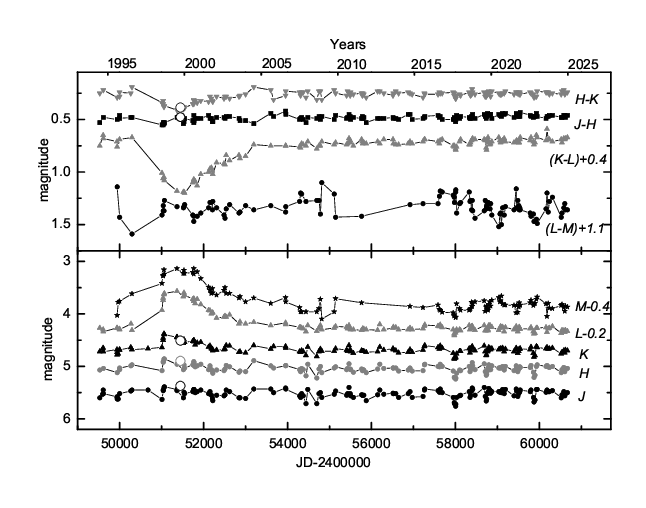}
    \caption{Near-IR light and color curves based on observations from 1995 to 2024. Open circles correspond to the data from the 2MASS catalog, obtained on September 27, 1999.}
    \label{fig:LC_IR}
\end{figure*}

In the near-IR range, particularly in the $KLM$ bands, the star’s variability exhibits a distinct pattern compared to the optical range. This is attributed to the presence of a circumstellar dust envelope, which significantly affects the object’s emission in the IR. Fig.~\ref{fig:LC_IR} presents the $JHKLM$ light and $J-H$, $H-K$, $K-L$, $L-M$ color curves. In the quiescent state, the amplitudes of quasi-periodic pulsation-related variations do not exceed: $\Delta J=0.^{m}25$, $\Delta H=0.^{m}25$, $\Delta K=0.^{m}20$, $\Delta L=0.^{m}10$. The larger amplitude of variability in the $M$ band ($\Delta M=0.^{m}35$) may be caused either by higher observational errors in this band compared to other near-IR wavelengths, or by possible instabilities in the contribution from the dust shell emission.

%----------------Fig. 5 -----------------------------

%--------------------------------------------------

\begin{figure*}
      \centering\includegraphics[scale=1.5]{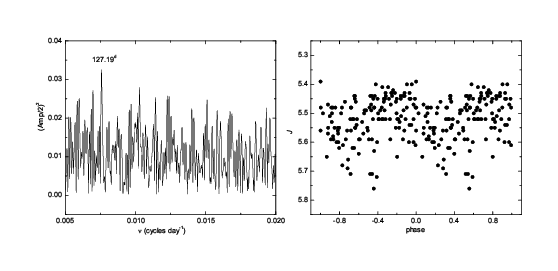}
    \caption{Periodogram and phase curve folded with a period of P=127.19\,d, based on the $J$-band dataset from 1995 to 2024.}
    \label{fig:phaseJ}
\end{figure*}

To search for periodic variations in the IR range, we used the $J$-band photometry, as it provided the most comprehensive observational coverage in our dataset and is less affected by the presence of a circumstellar dust envelope compared to other bands. The analysis revealed a period of 127.19\,d. After removing variations with this period, an additional period of 132.08\,d was detected. Both values are in good agreement with the results obtained from optical photometry by us and other authors. Fig.~\ref{fig:phaseJ} shows the periodogram based on $J$-band observations, along with the corresponding convolution using the dominant period of 127.19\,d.

%----------------Fig. 6 -----------------------------

%----------------------------------------------------

Of particular interest are the observations from 1996 to 2004 (August 22, 1996 -- July 13, 2004), when the star exhibited a brightness increase of 0.1, 0.7, and 0.7\,mag in the $K$, $L$, and $M$ bands, respectively, reaching a maximum by August 1999, followed by a gradual decline over five years back to the quiescent level. In the shorter-wavelength IR bands ($J$ and $H$) as well as in the optical range, where the star itself dominates the emission, the dust formation event had little effect. This can be explained either by a significant angle between the direction of the dust ejection and our line of sight, or by a low optical depth of the newly formed envelope. Otherwise, we would have observed a decrease in stellar brightness due to extinction by dust particles.

It is also worth noting the time delay in the maximum brightness between the $KLM$ bands during the mass-loss episode. The peak emission was first reached in the $K$ band, followed by the $L$ band, and finally in the $M$ band. This behavior is expected, as the dust cools while moving away from the stellar surface, causing its emission to shift toward longer wavelengths.

\begin{figure*}
   \centering   \includegraphics[scale=1.5]{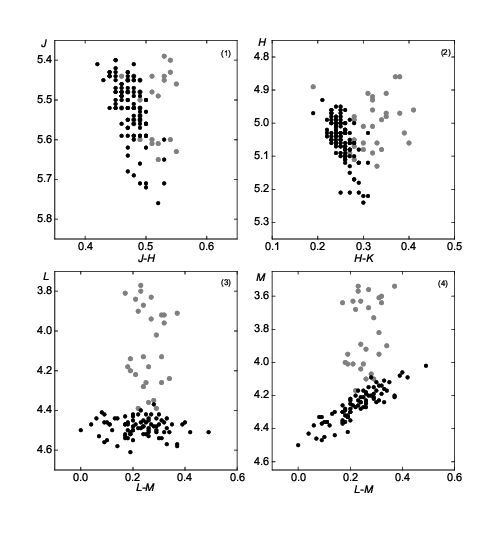}
    \caption{Color–magnitude diagrams for V354~Lac. Black dots mark the quiescent state; grey dots show data from 1996 to 2004, corresponding to the dust-formation episode.}
    \label{fig:CBIR}
\end{figure*}

Two distinct states of the star -- quiescence and the dust-formation episode -- are clearly separated in the color–magnitude diagrams shown in Fig.~\ref{fig:CBIR}. In the $J-(J-H)$ diagram (Fig.~\ref{fig:CBIR}a), only a slight reddening relative to the quiescent state is observed, while the brightness in the $J$-band remains nearly unchanged. A stronger reddening effect is seen in the $H-K$ color (Fig.~\ref{fig:CBIR}b), accompanied by a minor brightening in the $H$ band. In the $L$ and $M$ bands, the brightness variations plotted as a function of the $L-M$ color index (Figs.~\ref{fig:CBIR}c,d) reveal a significant increase in flux during the 1996--2004 event, while the color remains nearly constant. In the quiescent state, no correlation between $L$ magnitude and $L-M$ color is apparent, but a clear correlation between $L-M$ and $M$-band brightness is observed -- the object reddens as it brightens (correlation coefficient of 0.86). In contrast, in the optical range and in the $J$ and $H$ bands, an inverse trend is found -- the star becomes redder when fainter, which is consistent with temperature pulsations.

%----------------Fig. 7 -----------------------------

%--------------------------------------------------

\begin{figure*}
      \centering
      \includegraphics[scale=1.2]{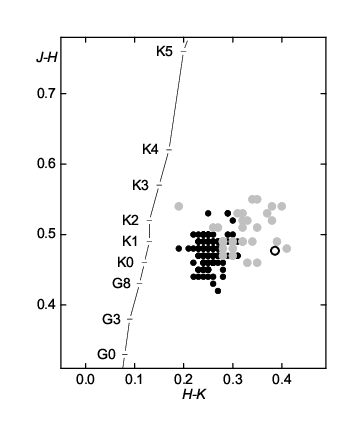}
    \caption{The $(H-K)$ vs. $(J-H)$ two-color diagram. Quiescent-state data are shown as black dots; grey dots (1996–2004) correspond to the dust ejection event. The open circle indicates the 2MASS observation from September 27, 1999.}

    \label{fig:2col-IR}
\end{figure*}

The two-color $(H-K)$ vs. $(J-H)$ diagram (Fig.~\ref{fig:2col-IR}) illustrates the location of V354~Lac in comparison to the sequence of intrinsic color indices of supergiants (Koornneef, 1983). It is evident that the color indices associated with the dust ejection episode are redder -- indicating an increased contribution from the dust component. The open symbol represents the 2MASS measurement from September 27, 1999 (Cutri et al., 2003) which is consistent with our photometric data.

%----------------Fig. 8 -----------------------------

%--------------------------------------------------

%--------------------------------------------------

\section*{DISCUSSION AND CONCLUSION}

Photometric analysis of the $UBV$ light curves and color indices confirms previous findings that V354~Lac exhibits variability with two closely spaced periods: 127 and 132\,d, which are attributed to stellar pulsations. Similar periodicities were also detected in the $J$-band data. The period ratio is $P_1/P_2=0.96$. This type of variability -- with closely spaced periods -- is observed in other F-G-type post-AGB stars, such as V1648~Aql ($P_1/P_2 = 0.96$) (Arkhipova et al., 2010), V448~Lac ($P_1/P_2=0.95$) (Hrivnak et al., 2013), IRAS~08544–4431 ($P_1/P_2 = 0.96$), and IRAS~12222–4652 ($P_1/P_2=0.91$) (Van Winckel et al., 2009). No significant changes in the pulsation periods have been detected over more than 30 years of $UBV$ photometric monitoring. In the $UBV$ light curves covering 2007--2024 (MJD=54400--60200), a long-term variation with a timescale of approximately 4000\,d appears to be present, though its significance remains uncertain at this stage of research. A new periodicity of 661\,d has been detected in the $UBV$ photometry, whose origin is currently unclear. Hrivnak et al. (2022) reported a similar period of $\sim$1.8\,yr and suggested that it could be associated with orbital motion in a binary system. Assuming a circular orbit with a period of 661\,d, and a primary mass of 0.61\,$M_\odot$ (Hrivnak et al. 2017), the orbital radius would be $\sim$340\,$R_\odot$ ($\sim$1.6\,AU), assuming negligible secondary mass. However, this result appears physically implausible in our cas, since the typical radius of an AGB star -- the progenitor of a proto-planetary nebula -- reaches 400-500\,$R_\odot$. Such a configuration would imply that the secondary component resides within the atmosphere of the primary, suggesting the need for further investigation. A similar issue was encountered by Hrivnak et al. (2017), who attempted to relate 700–800-d variations seen in radial velocity curves to orbital motion.

An important result of this study is the detection of a near-IR brightness increase at the turn of the century in the $KLM$ bands. The dust formation episode lasted approximately eight years (1996–2004), reaching its maximum around 1998-2001.

Earlier studies also reported episodes of enhanced mass loss in V354~Lac. Hrivnak et al. (1994) examined $K$-band spectra (2.13-2.53\,$\mu$m) from 1990-1991 and found that the CO spectrum transitioned from emission to absorption within three months, remaining in this state until at least November 4, 1991. They proposed that the CO emission originates in a shock wave produced by the interaction of a high-velocity outflow with the circumstellar envelope. The excitation of high-energy CO levels may be caused by a sudden mass ejection -- either at the base of the wind, where the density is highest, or in the region where the flow interacts with an outer dusty or molecular shell, leading to dissipation of the shock energy.

Ueta et al. (2001) obtained mid-IR images of the object at 8.8, 9.8, and 11.7\,$\mu$m on June 16, 2000, and at 7.9, 9.7, and 12.5\,$\mu$m on November 4, 2000. These observations correspond to the peak phase of the dust ejection event detected in our near-IR monitoring. The authors carried out radiative transfer modeling of the dust emission for IRAS~22272+5435, suggesting a structure composed of three distinct dust shells: a spherical envelope formed during the AGB phase; an inner shell associated with the superwind phase; and a third shell resulting from a dust ejection episode in the early 1990s. Notably, this last event coincides in time with the CO line variability reported by Hrivnak et al. (1994).

According to Ueta et al. (2001), the mass-loss rate during the AGB phase was $\dot{M} = 8 \times 10^{-7}\,M_\odot$/yr for a spherically symmetric wind and increased to $\dot{M} = 4 \times 10^{-6}\,M_\odot$/yr during the axisymmetric superwind phase. The authors concluded that V354~Lac left the AGB stage approximately 380\,years ago and experienced a sudden mass ejection about ten years prior to their observations -- around the early 1990s. Previous estimates of the nebula's dynamical age and mass-loss rates are inconsistent (see references in Ueta et al., 2001). The axisymmetric structure of the nebula is supported by multiple optical and infrared studies (Meixner et al., 1997; Dayal et al., 1998; Ueta et al., 2000, 2001).

The origin of episodic mass loss during the post-AGB phase is still not well understood. The observed rise in IR brightness with little change in optical magnitudes indicates that the dust envelope is optically thin at visible wavelengths and does not appreciably attenuate the stellar light.

The process of dust formation in the atmospheres of post-AGB stars and its relation to pulsational activity is discussed by Za\v cs et al. (2009). Pulsations give rise to waves that propagate into the upper atmospheric layers and evolve into shock waves. These shocks induce variations in temperature and density, which may trigger dust condensation. Radiative pressure on dust grains is considered one of the main drivers of stellar winds in AGB and post-AGB stars, leading to substantial mass-loss rates. While such mechanisms are known to operate in the atmospheres and circumstellar environments of pulsating late-type stars, the 1996-2004 dust episode observed in V354~Lac appears more pronounced than typical cases, suggesting the involvement of an additional mechanism contributing to dust formation.

As a possible explanation, the presence of a stellar companion to V354~Lac is suggested. In such a scenario, additional dust formation could occur in the wind-interaction zone between the two components. A similar mechanism is observed in Wolf-Rayet (WR) binary systems, such as WR~137 (HD~192641), a 13-yr-period system consisting of a carbon-rich WR star and an Oe-type companion. Dust production in this system is associated with the collision of their stellar winds, which creates regions of enhanced density where dust can form. The efficiency of dust formation increases near periastron, where the stars are closest and their wind velocities peak. The dust-formation events repeat with the orbital period (Peatt et al., 2023). Similar episodes of rapid IR brightening due to dust condensation have been previously reported in other WR binaries: WR~140 (HD~193793) (Williams et al., 1978) and WR~48a (Danks et al., 1983).

Unlike the WR-type binaries discussed above, V354~Lac exhibited only a single IR brightening event more than 20 years ago. This suggests that, if the star is indeed part of a binary system, its orbital period must exceed this timespan. Only long-term IR monitoring can reveal whether such an episode will recur. Notably, Hrivnak et al. (2017) presented radial velocity data indicating a possible variability period exceeding 30 years. A significant shift of 2.8\,km/s was found between mean velocities measured in recent (2007-2015) and early (1991-1995) observations. Assuming an 18-yr interval between these measurements, the orbital period must be at least 36\,yr. For such a period and a velocity semi-amplitude of 1.4\,km/s, with a primary mass of $M_1=0.62M_{\odot}$ and an orbital inclination of $i=25^{\circ}$, the secondary component would have a mass of $M_2=0.36M_{\odot}$ and orbit at a distance of approximately 11\,AU from the primary (Hrivnak et al., 2017).

For V354~Lac, a large body of observational data has been accumulated, significantly improving our understanding of the star’s nature and the mechanisms behind its variability. Nevertheless, several key questions remain unanswered: Is this object a binary system? Is the dust formation episode connected to the presence of a companion? Addressing these issues requires further investigation, primarily through new observations that can test predictions from theoretical models. The photometric data presented in this work will serve as a basis for future modeling of the circumstellar dust shell in two distinct phases: quiescence and the 1996-2004 dust ejection event.

\bigskip
\section*{Data Availability Statement} The photometric data reported in this study are available on request to the first author via email at ikonnikova@sai.msu.ru.

\section*{Acknowledgement}

The study was conducted under the state assignment of Lomonosov Moscow State University.

\bigskip
REFERENCES
\begin{enumerate}

\item V.P.~Arkhipova, N.P.~Ikonnikova, and R.I.~Noskova, Astron. Lett. {\bf 19},  (1993).

\item V.P.~Arkhipova, N.P.~Ikonnikova, R.I.~Noskova, and G.V.~Sokol, Astron. Lett. {\bf 29},  (2000).

\item V.P.~Arkhipova, N.P.~Ikonnikova, and G.V.~Komissarova, Peremennye Zvezdy, {\bf 29}, 1 (2009).

\item V.P.~Arkhipova, N.P.~Ikonnikova, and G.V.~Komissarova, Astron. Lett. {\bf 36},  (2010).

\item V.P.~Arkhipova, O.G.~Taranova, N.P.~Ikonnikova, V.F.~Esipov, G.V.~Komissarova, and V.I.~Shenavrin, Astron. Lett. {\bf 42},  (2016).

\item V.~Bujarrabal, A.~Castro-Carrizo, J.~Alcolea, and C.~S\'{a}nchez Contreras, Astron. Astrophys. {\bf 377}, 868 (2001).

\item P.~Cox, Astron. Astrophys. {\bf 236}, L29 (1990).

\item R.M.~Cutri, M.F.~Skrutskie, S.~van Dyk, et al., VizieR Online Data Catalog, II/328 (2003).

\item A.C.~Danks, M.~Dennefeld, W.~Wamsteker, and P.A.~Shaver, Astron. Astrophys. {\bf 118}, 301 (1983).

\item A.~Dayal, W.~Hoffmann, J.H.~Bieing, et al., Astrophys. J. {\bf 392}, 603 (1998).

\item T.J.~Deeming, Astrophys. Space Sci. {\bf 36}, 137 (1975).

\item K.~De Smedt, H.~Van Winckel, D.~Kamath, L.~Siess, S.~Goriely, A.I.~Karakas, and R.~Manick, Astron. Astrophys. {\bf 587}, A6  (2016).

\item D.~Fong, M.~Meixner, E.C.~Sutton, A.~Zalucha, and W.J.~Welch, Astrophys. J. {\bf 652}, 1626 (2006).

\item J.H.~Goebel, Astron. Astrophys. {\bf 278}, 226 (1993).

\item B.J.~Hrivnak and S.~Kwok, Astrophys. J. {\bf 371}, 631 (1991).

\item B.J.~Hrivnak, S.~Kwok, and T.R.~Geballe, Astrophys. J. {\bf 420}, 783 (1994).

\item B.J.~Hrivnak and W.~Lu,  {\it The Carbon Star Phenomenon. IAU Symp. {\bf 177}} (Ed. R.F.~Wing, Dordrecht: Kluwer Acad. Publ., 2000), p. 293.

\item B.J.~Hrivnak and J.H.~Bieging, Astrophys. J. {\bf 624} 624, 331 (2005).

\item B.J.~Hrivnak, W.~Lu, R.E.~Maupin, and B.D.~Spitzbart, Astrophys. J. {\bf 709}, 1042 (2010).

\item B.J.~Hrivnak, W.~Lu, J.~Sperauskas, H.~Van Winckel, D.~Bohlender, and L.~Za\v{c}s, Astrophys. J. {\bf 766}, 116 (2013).

\item B.J.~Hrivnak, G.~Van de Steene, H.~Van Winckel, J.~Sperauskas, D.~Bohlender, and W.~Lu, Astrophys. J. {\bf 846}, 96 (2017).

\item B.J.~Hrivnak, W.~Lu, W.C.~Bakke, and P.J.~Grimm, Astrophys. J. {\bf 939}, 32 (2022).

\item H.L.~Johnson, R.I.~Mitchel, B.~Iriarte, and W.Z.~Wisniewski, Comm. Lunar and Planet. Lab. {\bf 4}, 99 (1966).

\item V.G.~Klochkova, V.E.~Panchuk, N.S.~Tavolganskaya, Astrophys. Bull. {\bf 64}, 155 (2009).

\item J.~Koornneef, Astron. Astrophys. {\bf 128}, 84 (1983).

\item S.~Kwok,  K.~Volk, and B.J.~Hrivnak, Astrophys. J. {\bf 345}, 51 (1989).

\item M.~Lindqvist, L.A.~Nyman, H.~Olofsson, and A.~Winnberg, Astron. Astrophys. {\bf 205}, L15 (1988).

\item V.M.~Lyutyi, Soobshch. GAISh 172, 30 (1971)

\item M.~Meixner, C.J.~Skinner, J.R.~Graham, E.~Keto, J.G.~Jernigan, and J.F.~Arens, Astrophys. J. {\bf 482}, 897 (1997).

\item J.~Nakashima, N.~Koning, N.H.~Volgenau, S.~Kwok, B.H.K.~Yung, and Y.~Zhang,  Astrophys. J. {\bf 759}, 61 (2012).

\item A.~Omont, H.S.~Moseley, P.~Cox, W.~Glaccum, S.~Casey, et al., Astrophys. J. {\bf 454}, 819 (1995).

\item M.J.~Peatt, N.D.~Richardson, P.M.~Williams, N.~Karnath, V.I.~Shenavrin, R.M.~Lau, A.F.J.~Moffat, and G.~Weigelt), Astrophys. J. {\bf 956}, 109 (2023).

\item Th.~Posch, H.~Mutschke, and A.~Andersen, Astrophys. J. {\bf 616}, 1167 (2004).

\item S.R.~Pottasch and M.~Parthasarathy, Astron. Astrophys. {\bf 192}, 182 (1988).

\item B.E.~Reddy, D.L.~Lambert, G.~Gonzalez, and D.~Yong, Astrophys. J. {\bf 564}, 482 (2002).

\item N.N.~Samus, E.V.~Kazarovets, O.V.~Durlevich, N.N.~Kireeva, and E.N.~Pastukhova, Astron. Rep. {\bf 61}, 80 (2017). 

\item V.I.~Shenavrin, O.G.~Taranova, A.E.~Nadzhip, Astron. Rep. {\bf 55}, 31 (2011).

\item V.~Strai\v zys, {\it Multicolor stellar photometry. Photometric systems and methods} (Vilnius: Mokslas Publishers, 1977).

\item W.~Strohmeier and R.~Knigge, {Veroeffentlichungen der Remeis-Sternwarte zu Bamberg} {\bf 5}, Nr.5, 1 (1960).

\item R.~Szczerba, N.~Siodmiak, G.~Stasi\'{n}ska, and J.~Borkowski, Astron. Astrophys. {\bf 469}, 799 (2007).

\item T.~Ueta, M.~Meixner, and M.~Bobrowsky, Astrophys. J. {\bf 528}, 861 (2000).

\item T.~Ueta, M.~Meixner, P.M.~Hinz, W.F.~Hoffmann, W.~Brandner, A.~Dayal, L.K.~Deutsch, G.G.~Fazio, and J.L.~Hora, Astrophys. J. {\bf 557}, 831 (2001).

\item W.E.C.J.~van der Veen, H.J.~Habing, and T.R.~Geballe, Astron. Astrophys. {\bf 226}, 108 (1989).

\item H.~Van Winckel, T.~Lloyd Evans, M.~Briquet, et al., Astron. Astrophys. {\bf 505}, 1221 (2009).

\item P.M.~Williams, D.H.~Beattie, T.J.~Lee, J.M.~Stewart, and E.~Antonopoulou, MNRAS {\bf 185}, 467 (1978).

\item L.~Za\v{c}s, V.G.~Klochkova, and V.E.~Panchuk, MNRAS {\bf 275}, 764 (1995).

\item L.~Za\v{c}s, M.R.~Schmidt, and R.~Szczerba, MNRAS 306, 903 (1999).

\item L.~Za\v{c}s, J.~Sperauskas, F.A.~Musaev, O.~Smirnova, T.C.~Yang, W.P.~Chen, and M.~Schmidt, Astrophys. J. {\bf 695}, L203 (2009).

\item L.~Za\v{c}s, F.A.~Musaev, B.~Kaminsky, Y.~Pavlenko, A.~Grankina, J.~Sperauskas, and B.J.~Hrivnak, Astrophys. J. {\bf 813}, id. 3 (2016).

\item L.~Za\v{c}s and K.~Pu\c{k}\={i}tis, Astrophys. J. {\bf 920}, 17 (2021).

\item K.~Zhang, B.W.~Jiang, and A.~Li,  MNRAS {\bf 396}, 1247 (2009).

\item B.~Zuckerman, H.M.~Dyck, and M.J.~Claussen, Astrophys. J. {\bf 304}, 401 (1986).

\end{enumerate}
\end{document}